\documentclass[a4paper,10pt]{article}
\usepackage{graphicx}
\begin{document}

\begin{titlepage}

\begin{center}
 \Large{\bf{Long-range interacting classical systems:\\
 universality in mixing weakening}}
\end{center}
\vspace{5mm}

\begin{center}
 \large{Alessandro Campa$^{1, \ddag}$, Andrea Giansanti$^{2, \S}$,
 Daniele Moroni$^{2, \dag}$ and Constantino Tsallis$^{3,4,5,*}$}
\end{center}
\vspace{5mm}

\begin{center}
 \normalsize{$^1$Laboratorio di Fisica, Istituto Superiore di Sanit\`a\\ and
 INFN Sezione di Roma1, Gruppo Collegato Sanit\`a\\Viale Regina
 Elena 299, 00161 Roma, Italy}
\end{center}
\begin{center}
 \normalsize{$^2$Dipartimento di Fisica, Universit\`a di
 Roma ``La Sapienza''\\ and INFM Unit\`a di Roma1, \\
 Piazzale Aldo Moro 2, 00185 Roma, Italy}
\end{center}
\begin{center}
 \normalsize{$^3$Centro Brasileiro de Pesquisas F\'{\i}sicas\\
 Rua Xavier Sigaud 150, 22290 -180 Rio de Janeiro - RJ, Brazil}
\end{center}
\begin{center}
 \normalsize{$^4$The Santa Fe Institute, \\1399 Hyde Park Road,
 Santa Fe, New Mexico 87501, USA}
\end{center}
\begin{center}
 \normalsize{$^5$Department of Mechanical Engineering\\
 Massachusetts Institute of Technology\\ 77 Massachusetts Avenue,
 Cambridge, Massachusetts 02139, USA}
\end{center}
\vspace{5mm}
\begin{center}
\large{\bf {Abstract}}
\end{center}
\vspace{2mm}
Through molecular dynamics, we study the $d=2,3$ classical model of
$N$ coupled rotators (inertial $XY$ model) assuming a coupling
constant which decays with distance as $r_{ij}^{-\alpha}$ ($\alpha
\ge 0$). The total energy $<H>$ is asymptotically $\propto N {\tilde N}$
with ${\tilde N} \equiv [N^{1-\alpha/d}-(\alpha/d)]/[1-\alpha/d]$,
hence the model is thermodynamically extensive if $\alpha/d>1$ and
nonextensive otherwise. We numerically show that, for energies above
some threshold, the (appropriately scaled) maximum Lyapunov exponent
is $\propto N^{-\kappa}$ where $\kappa$ is an {\it universal} (one
and the same for $d=1,2$ and $3$, and all energies) function of
$\alpha/d$, which monotonically decreases from 1/3 to zero when
$\alpha/d$ increases from 0 to 1, and identically vanishes above 1.
These features are consistent with the nonextensive statistical
mechanics scenario, where thermodynamic extensivity is associated
with {\it exponential} mixing in phase space, whereas {\it weaker}
(possibly {\it power-law} in the present case) mixing emerges at the
$N \rightarrow \infty$ limit whenever nonextensivity is observed.
\vspace{5mm}

\vspace{5mm}

\noindent$^{\S}$\small{Corresponding author; Andrea.Giansanti@roma1.infn.it}\\
$^{\ddag}$\small{Campa@iss.infn.it}\\
$^{\dag}$\small{smoroni@lucifero.phys.uniroma1.it}\\
$^{*}$\small{tsallis@cbpf.br}

\end{titlepage}

In the last few years there has been a noticeable interest in the
study of the thermodynamics and statistical mechanics of anomalous
systems exhibiting nonextensivity \cite{T1,T2}. A system is extensive
if its energy and entropy, as functions of intensive internal parameters 
(e.g. temperature), grow linearly with the size of the system
(i.e. with $N$, the number of its microscopic components).
Nonextensivity can be brought into scene by long-range interactions.
From the static point of view, nonextensivity means that
thermodynamic quantities like the internal or the free energies {\it
per particle}  are not constant with $N$, but rather
diverge in the thermodynamic limit $N\rightarrow\infty$. From the
dynamical point of view, one might think that long-range interactions 
induce persistent
spatial and temporal correlations leading to the breaking of standard
mixing and ergodicity properties, hence to a possible violation of
the usual Boltzmann-Gibbs (BG) statistics. Classical Hamiltonian
systems with many degrees of freedom are paradigmatic in the
discussion of the foundations of equilibrium statistical mechanics.
The maximal Lyapunov
exponent (MLE) is a well established indicator of chaos. 
If it is positive, the system will generically be strongly
chaotic and will satisfy the standard ergodic hypothesis attained
through {\it exponentially} quick mixing. If it is instead zero
non ergodicity could emerge, giving
origin to less-than-exponential (typically {\it power-law}) mixing,
hence to anomalous thermostatistical behavior. It is then interesting
to study maximal Lyapunov exponents in systems with long-range
interactions, where nonextensive thermodynamical and dynamical
anomalies are expected. In a previous work \cite{AT} the maximal
Lyapunov exponent of a one-dimensional system of $N$ planar rotators
coupled with interactions decaying as the inverse power $\alpha$ of
their distances (see model (\ref{ham1}) below) has been studied as a
function of $N$. For total energy above some threshold, the maximal
Lyapunov exponent is proportional to $1/N^{\kappa}$, where
$\kappa(\alpha)$ is a function which goes from $\kappa(0)>0$ to zero
while $\alpha$ increases from zero to 1 (long range forces) and
remains zero for $\alpha > 1$ (short range forces). In this letter we
want to investigate if the exponent $\kappa$, describing the
weakening of the mixing properties of the dynamics in the
long-range-interacting regime, is universal. In particular, we check
the hypothesis that $\kappa(\alpha,d)$ is a universal function of
$\alpha/d$, where $d$ is the dimensionality of space.

Let us consider a simple model of $N$ planar rotators (XY spins)
placed at the sites $i=1\ldots N$ of a square and a cubic lattice
(i.e., dimension $d$ = 2 and 3). The rotators have unit moment of
inertia, angular momentum $L_i$ and position specified by the angle
$\theta_i\in [0,2\pi]$. Rotations occur inside an arbitrary 
reference plane. The
model is described by the following classical hamiltonian in the
conjugate canonical pairs $\{\theta_i,L_i\}$:
\begin{equation}
\label{ham1} H =K+V =\frac{1}{2} \sum_i L_i^2 \,
+ \, \frac{1}{2} \sum_{ij,i\neq j} \frac {1-\cos(\theta_i-\theta_j)}
{r_{ij}^\alpha};
\end{equation}
where $r_{ij}$ is the distance between lattice site $i$ and $j$ with
periodic boundary conditions in the form of nearest image convention;
$\alpha \geq 0$ is a variable parameter specifying the range of the
interaction, which is defined long-range if $\alpha \le d$ and 
short-range if $\alpha>d$.

This system is nonextensive in the case of  long-range interactions
\cite{fisheretal}. To verify that, let us consider as an example a
quantity related to the internal energy per rotator at $T=0$ (i. e.,
when $K=0$ and all rotators are equally oriented), namely the
integral $\int_1^{\infty} dr r^{d-1} r^{-\alpha}$, which converges
for $\alpha > d$ and diverges for $0 \leq \alpha \leq d$. In the
latter case, the thermodynamics of the model is not well defined in
the usual sense. However, it is possible to define energy quantities
that are bounded in the thermodynamic limit if they are properly
scaled (see \cite{T2,tsallisfractals,ST}). It is interesting to
remark that, in system (\ref{ham1}) for $0 \leq \alpha \leq d$, not
only equilibrium (static) quantities, but also a quantity related to
the local instability of the dynamics such as the maximal Lyapunov
exponent might diverge with $N$ \cite{note1}, and needs to be
properly rescaled by a
function of $N$ if one wants to evaluate the degree of chaoticity as
a function of the size of the system.

An alternative procedure is to force the system to be extensive via a
largely used mean-field-like rescaling \cite{note2}
of the potential energy in (\ref{ham1}) which then becomes:
\begin{equation}
\label{ham2} H'=K'+V'=\frac{1}{2} \sum_i L_i^2 \, + \,
\frac{1}{2\tilde{N}} \sum_{ij,i\neq j} \frac
{1-\cos(\theta_i-\theta_j)} {r_{ij}^\alpha}
\end{equation}

The rescaling factor $\tilde{N}$ in front of the potential energy $V$
of hamiltonian (\ref{ham2}), is a function of $\alpha,d,N$ and of the
geometry of the lattice and has to be chosen to guarantee the
existence of a bounded energy density $U=<H'/N>$ in the thermodynamic
limit.

\begin{figure}[htbp]
\begin{center}
\includegraphics[width=9cm,bb=50 50 554 554,keepaspectratio,
angle=-90]{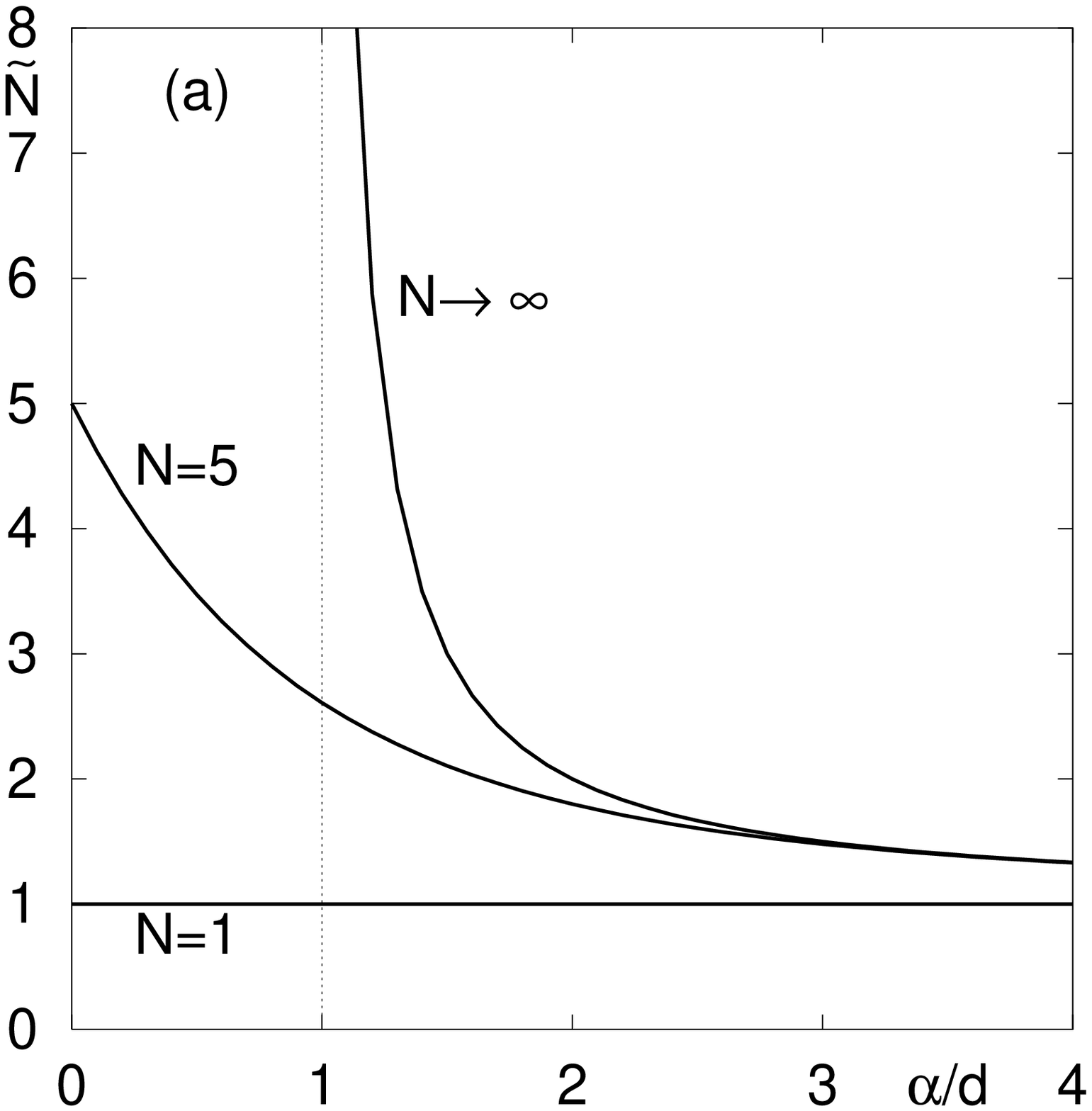}
\includegraphics[width=9cm,bb=50 50 554 554,keepaspectratio,
angle=-90]{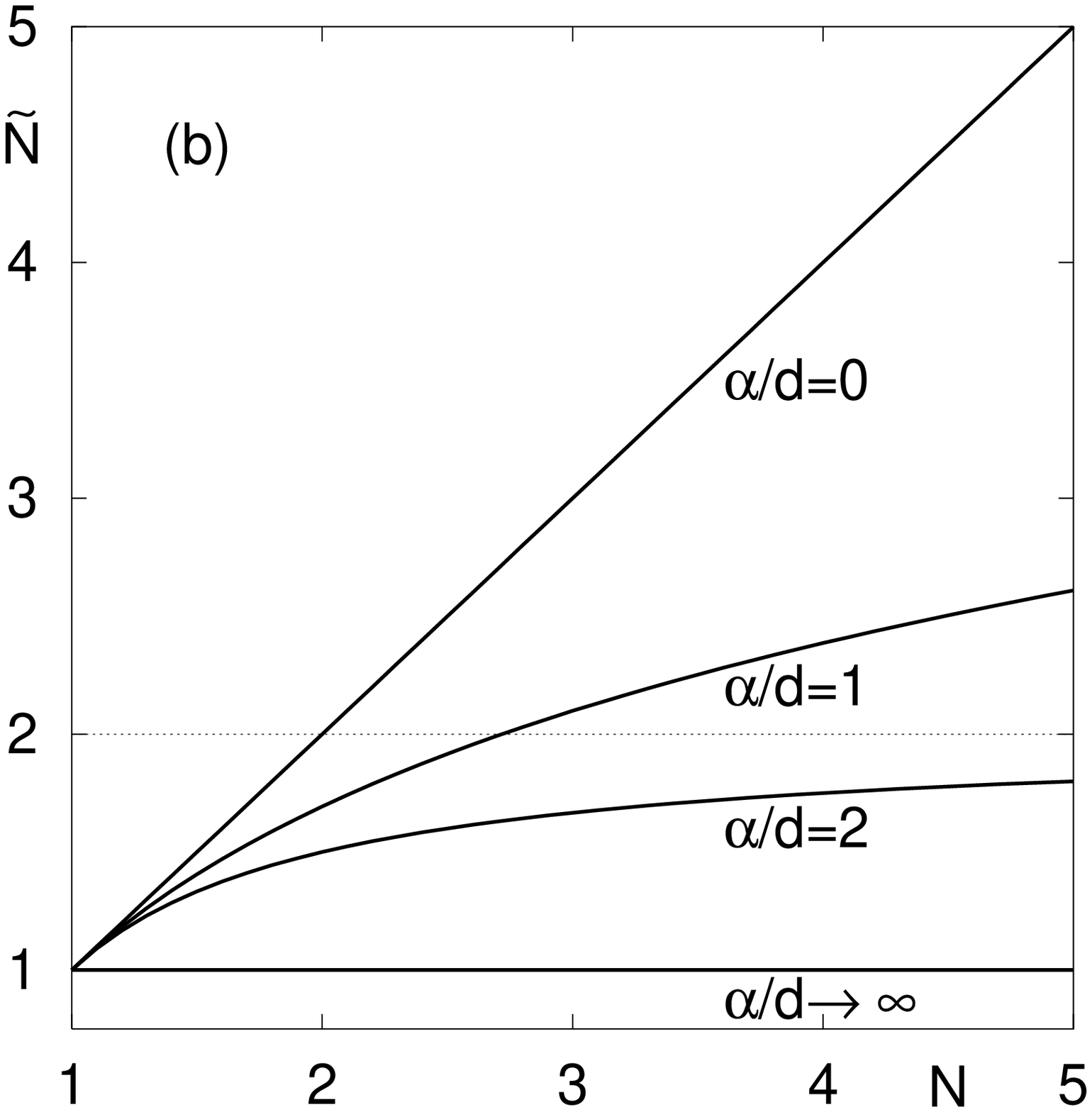}
\end{center}
\caption{\small Scaling factor $\tilde{N}(N,\alpha/d)$: (a) versus
$\alpha/d$, for typical values of $N$ (dotted line: the $N
\rightarrow \infty$ asymptote); (b) versus $N$, for typical values of
$\alpha/d$ (dotted line: the $\alpha/d=2$ asymptote). Note the divergence
of $\tilde{N}$ in the thermodynamic limit ($N\rightarrow\infty$), in the
long-range case.}
\label{Ntilde1}
\end{figure}

The model (1) has been introduced in \cite{AT} on a one-dimensional
lattice and its thermodynamics has been studied by means of
simulations in \cite{TA}. When $\alpha=0$ and $\tilde{N}=N$ the model
is also called the Hamiltonian Mean-Field XY Model (HMF) \cite{AR},
whose dynamics and thermodynamics have been extensively studied
\cite{AnT,LRR0,LRR}. In this case the underlying lattice has no
meaning and any dependence from its dimension $d$ disappears. It is
found that the HMF model has a phase transition from a ferromagnetic
state to a paramagnetic one at the critical temperature $T_c=0.5$
corresponding to a critical energy density $U_c=0.75$ ($k_{B}=1$). In
this letter we use \cite{TBJP}
\begin{eqnarray}
&& \tilde{N} \equiv1 + d \int_1^{N^{1/d}} dr r^{d-1-\alpha} =
\left \{
\begin{array}{rl}
\frac{N^{1-\alpha/d}-\alpha/d}{1-\alpha/d} & \mbox{if } \alpha \neq d \\
1 + \ln N & \mbox{if } \alpha = d
\end{array}
\right. \label{nstar}
\end{eqnarray}
whose properties are shown in figure \ref{Ntilde1}. 
(Unity has been added
to the integral in the definition of $\tilde{N}$ in order to have the
convenient limit $\tilde{N} \rightarrow 1$ when $\alpha/d \rightarrow
\infty$.) Using a similar rescaling factor it was shown in \cite{CGM}
that model (\ref{ham2}) and HMF model have the same thermodynamics,
and in particular the same critical energy density.

Chaoticity properties and maximal Lyapunov exponent (MLE) of HMF model have
been studied by Latora, Rapisarda and Ruffo
\cite{LRR,LRR2,LRR3,LRR4}. They calculated the curve
MLE versus $U$
numerically for various $N$ and analytically for
$N\rightarrow\infty$; MLE goes to 0 both when $U$ tends
to $0$ and when $U$ tends to $\infty$. In these two limits the
system becomes integrable, being represented by a set of harmonic
oscillators around the ferromagnetic fundamental state of aligned
rotators in the former case and by free rotators in the latter.
Finite-size effects are present for $U>U_c$ where MLE
vanishes in the thermodynamic limit as $\sim N^{-1/3}$. For fixed
$U\leq U_c$ the curve is smooth and presents a positive maximum at
the critical energy $U_c$, where therefore in the thermodynamic limit
exists a finite discontinuity.\cite{MCF}

The behavior of MLE in the one-dimensional model
(\ref{ham2}) has been studied, as said above, in \cite{AT}, as a
function of $N$ and $\alpha$ at a fixed high energy (above $U_c$). It
has been shown that it goes to 0 in the thermodynamic
limit when $\alpha\leq d$, like in the HMF model, but it remains
constant when $\alpha> d$. The way the exponent vanishes is still of
the form $\sim N^{-\kappa}$ where now $\kappa$ is a function of
$\alpha$. The curve $\kappa(\alpha)$ starts from the value $1/3$ at
$\alpha=0$ and goes to 0 when $\alpha\rightarrow 1^-$.

In the present work we extend this study to  cases $d>1$. We still
expect that when $\alpha <
d$ and $U>U_c$ MLE vanishes in the 
thermodynamic limit with a power law of the type
$N^{-\kappa}$ with  $\kappa=\kappa(\alpha,d)$ satisfying
$\kappa(\alpha=0,d)=1/3$ \cite{MCF} and vanishing when
$\alpha\rightarrow d^-$. Two questions seem quite relevant. Is this
curve $\kappa(\alpha,d)$ universal  in the sense that it {\it only}
depends on $\alpha/d$, or it also depends  on the dimensionality $d$
of the system ? Moreover, is this curve  independent from the energy
density $U$ at which it is computed? In what follows we will
positively answer to both questions.

We have simulated the constant energy dynamics of the nonextensive
hamiltonian (\ref{ham1}) and computed the maximal Lyapunov exponents.

Hamiltonians $H$ (eq. (\ref{ham1})) and $H'$  (eq. (\ref{ham2})) are
strictly connected. If we multiply $H'$ by $\tilde{N}$, the potential
energy $V'$ becomes $V$. Then, if we rescale time as
$t'=t\sqrt{\tilde{N}}$, and using the fact that $L_i$'s involve first
order time derivatives, we obtain that also the kinetic energy $K'$
becomes $K$. Finally we get $H=\tilde{N}H'$.

Integrating the equations of motion of $H$ is identical to
integrating those of $\tilde{N}H'$  but for times longer by a factor
$\sqrt{\tilde{N}}$. All thermodynamic quantities that can be obtained
as temporal averages of time dependent observables are not influenced
by the rescaling of time. In particular, once one has $T,V,U$ from
$H$ simulations, one can get the corresponding values for $H'$ simply
dividing by $\tilde{N}$. The evaluation of maximal Lyapunov exponents
$\lambda_{max}'$ of hamiltonian $H'$ can also be obtained through
rescaling of the one measured for $H$, noted $\lambda_{max}$
hereafter.

In order to calculate maximal Lyapunov exponents one must consider
the limit, \cite{BGS}
\begin{equation}\label{lambdat}
\lambda_{max}=\lim_{t\rightarrow\infty}\frac{1}{t}\ln
\frac{d(t)}{d(0)}=\lim_{t\rightarrow\infty}\lambda(t)
\end{equation}
with $d(t)=\sqrt{\sum_i (\delta \theta_i)^2 + (\delta L_i)^2}$ being
the metric distance calculated from infinitesimal displacements at
time $t$ obtained in turn through the double integration of both the
normal equations of motion
\begin{eqnarray}
 \dot{\theta_i} &=& L_i \\ \label{eqmoto.a}
 \dot{L_i}&=&\sum_{j,j\neq i}
 \frac{\sin(\theta_j-\theta_i)}{r_{ij}^\alpha} \label{eqmoto.b}
\end{eqnarray}
and the linearized ones
\begin{eqnarray}
\dot{\delta\theta_i}&=&\delta L_i \label{eqmototg.a} \\
\dot{\delta L_i}&=& \sum_{j\neq
i}\frac{\cos(\theta_j-\theta_i)}{r_{ij}^\alpha}
(\delta\theta_j-\delta\theta_i)\label{eqmototg.b}
\end{eqnarray}
(we report the ones derived from $H$, the ones for $H'$ are exactly
the same with the substitution ${r_{ij}^\alpha} \rightarrow
{r_{ij}^\alpha}\tilde{N}$). Because of formula (\ref{lambdat}) which
involves $1/t$, time rescaling yields
$\lambda_{max}'={\lambda_{max}}/{\sqrt{\tilde{N}}}$.

We computed $\lambda_{max}$ for typical values of $N\simeq 30$ to
$4000$ and a fixed energy density $U=5.0$ above the critical one
$U_c=0.75$. We made simulations on a square and simple cubic lattices
(i.e., $d=2,3$) with unitary lattice step, for various values of the
$\alpha/d$ ratio.

\begin{figure}[htbp]
\begin{center}
\includegraphics[width=\textwidth,height=15cm]{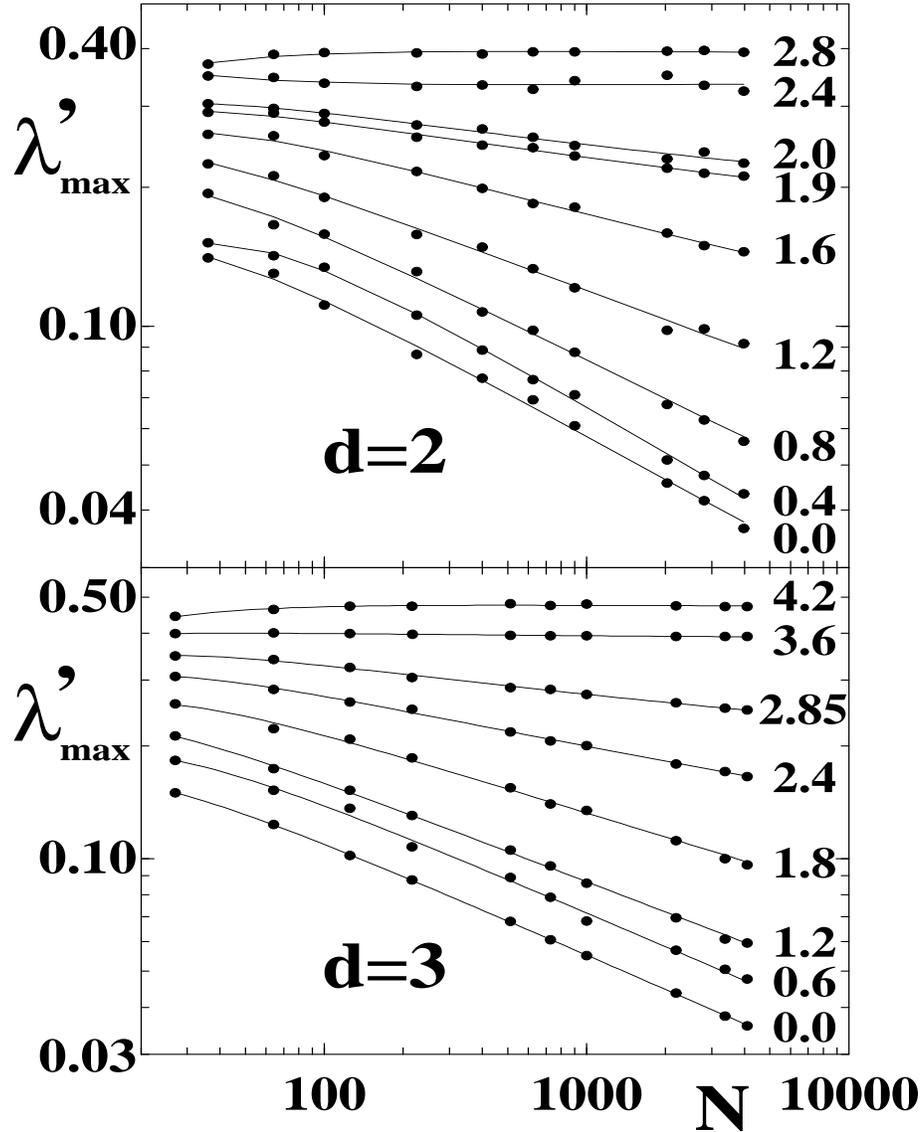}
\end{center}
\caption{\small Curves $\lambda_{max}'$ vs $N$ (log-log
  scale) for $d=2$ (upper part) and $d=3$ (lower part), $U=5$.  Solid
  lines are fits with the functional form
  $\left(a-\frac{b}{N}\right)/(\tilde{N})^c$. Consequently,
  $\lambda_{max}' \propto N^{-\kappa(\alpha,d)}$ where
  $\kappa(\alpha,d)= (1-\alpha/d)c(\alpha,d)$ for $0\leq\alpha < d$ and
  $\kappa(\alpha,d)=0$ for $\alpha>d$; for $\alpha=d$,
  $\lambda_{max}'$ is expected to vanish as a power of $1/\ln N$.
  The ordinate indistinctively represents $\lambda_{max}'$ corresponding to
  Hamiltonian (2) with $U=H'/N=5$, or $\lambda_{max}/\sqrt{\tilde{N}}$
  corresponding to Hamiltonian (1) with $U=H/(N\tilde{N})=5$.}
\label{prliap1}
\end{figure}

Simulations started from an high temperature initial state:
$\theta_i$'s are randomly extracted from a uniform distribution in
$[0,2\pi]$, momenta $L_i$'s are extracted from a uniform distribution
in $[-0.5,0.5]$ and then translated and rescaled to have zero total
momentum and the desired total energy. We  used the velocity-Verlet
algorithm \cite{SABW} with  a time-step chosen to have a relative
energy conservation of $10^{-4}$ or better. Length of simulations
were chosen looking at the asymptotic behavior of quantity
$\lambda(t)$, eq. (\ref{lambdat}), where $d(0)$ is randomly chosen
(see \cite{BGS}). We report in figure \ref{prliap1},
 the curves $\lambda_{max}'(N)$ for various $\alpha$ in
dimensions $d=2$ and $d=3$.

These curves extend those of fig. 3 in \cite{AT}. In both cases we
observe that, for growing $N$, if $\alpha>d$ then $\lambda_{max}'$ is
positive and constant, whereas if $\alpha<d$ the maximal Lyapunov
exponent tends to zero.

We fitted the data with the following functional form:
${\lambda}_{max}'\propto N^{-\kappa}$; $\kappa$ being the slope in
the log-log plots of fig. \ref{prliap1}.

\begin{figure}[ht]
\begin{center}
\includegraphics[width=\textwidth,keepaspectratio]{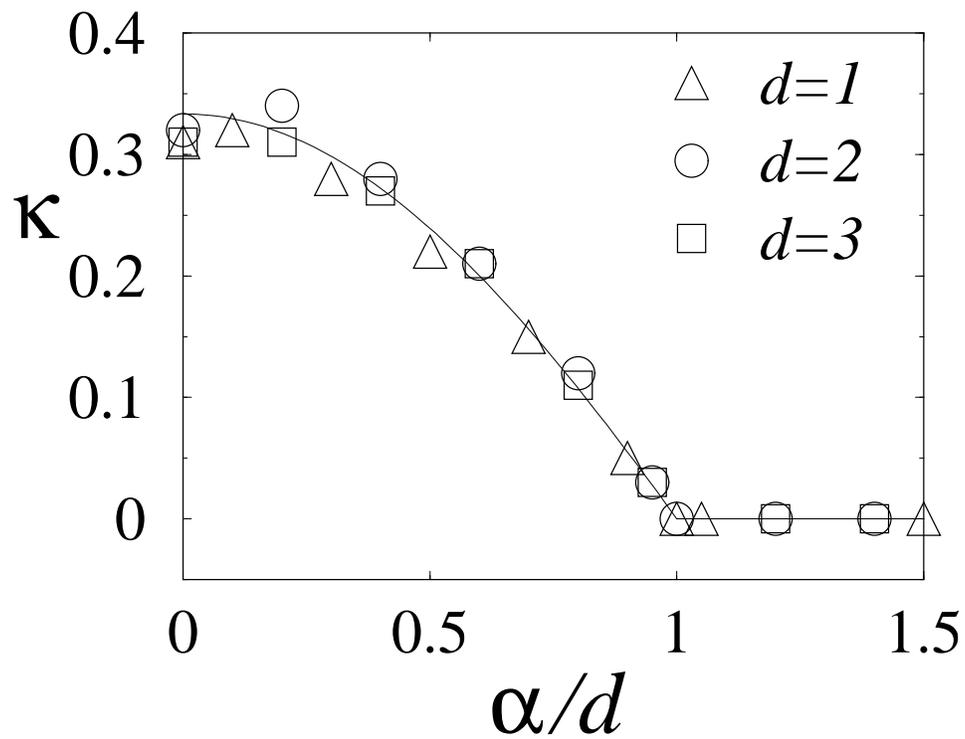}
\end{center}
\caption{\small The mixing weakening exponent $\kappa$ vs $\alpha/d$
for $d=1,2,3$ ($d=1$: from \cite{AT}; $d=2,3$: present work).
It describes the asymptotic $N$ behaviour of the maximal Lyapunov exponent
$\lambda_{max}'$ at fixed energy above the critical one, i.e.,
$\lambda_{max}' \propto 1/N^{\kappa}$. The solid line is a guide to the eye
consistent with universality. For $\alpha=0$ we have
\cite{MCF} $\kappa(0)=1/3 \; (\forall d)$.}
\label{prliap2}
\end{figure}

In figure \ref{prliap2} we collect the slopes $\kappa$ as a function
of the ratio $\alpha/d$, for $d=1$ (data from \cite{AT}), $d=2$ and
$d=3$. Remarkably, through the simple scaling $\alpha/d$ we show that
the $N$ dependence of the rescaled maximal Lyapunov exponent of model
(\ref{ham1}) is universal. All curves start at $\alpha=0$ from a
value close to $1/3$, analytically and numerically found for the HMF
model \cite{MCF,LRR2}; when $\alpha/d$ increases from zero to unity
all our data collapse into a single curve, and then remain zero for
$\alpha/d$ greater than unity. All of our data are roughly fitted by
the heuristic expression $\kappa=
[1-(\alpha/d)^2]/[3+(\alpha/d)^2/2]$.

Extrapolating for $N$ going to infinity  the results in fig.
\ref{prliap1}, it is interesting to note that the constant $\kappa$
behavior for $\alpha>d$ is consistent with the results found for the
$\alpha=\infty$ (only first-neighboring coupling) model \cite{BC}.

\begin{figure}[hb]
\begin{center}
\includegraphics[width=\textwidth,keepaspectratio]{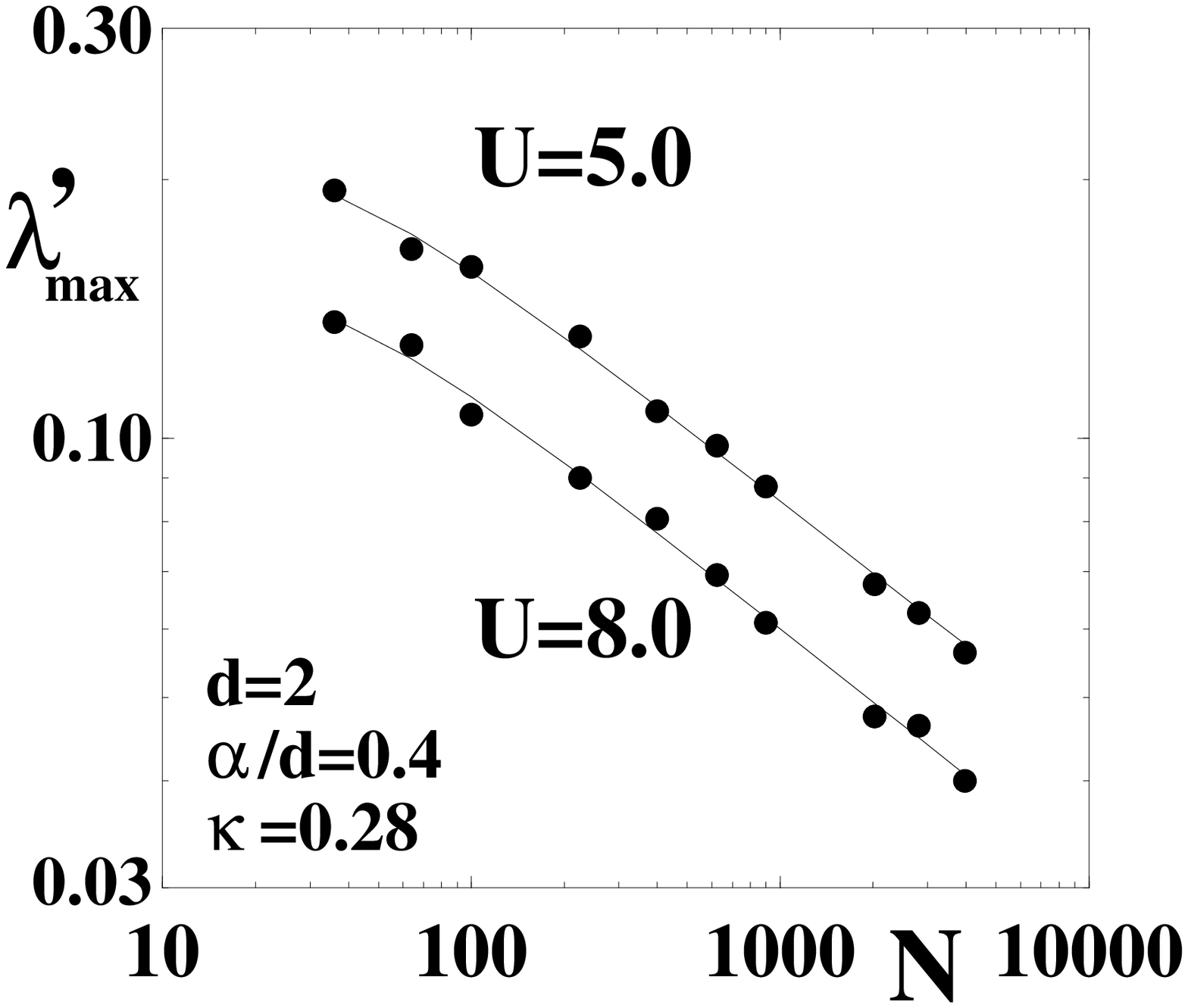}
\end{center}
\caption{\small $\lambda_{max}'(N)$ for fixed $\alpha=0.8$, $d=2$ and
two different energies $U$. The asymptotic $N$ behavior, for all
values of $U$ above the critical one $U_c(\alpha,d)$ and $0 \le
\alpha/d <1$, appears to be $\lambda_{max}' \propto
1/N^{\kappa(\alpha/d)}$, where the proportionality coefficient
decreases from a finite $(\alpha,d)$-dependent positive value to zero
when $U$ increases from $U_c(\alpha,d)$ to infinity.}
\label{prliap3}
\end{figure}

The universal function $\kappa$ shown in fig. \ref{prliap2} does not
appear to depend on the energy density, provided this is greater than
the critical one. For the case $\alpha=0$ this has been proved
analytically in \cite{MCF} and numerically in \cite{LRR2}. We show in
fig. \ref{prliap3} that this is also true for $\alpha=0.8$ and $d=2$,
the asymptotic slope of $\lambda_{max}'$ being sensibly the same for
the two energy densities $U=5$, $U=8$.

Summarizing our results, we can say that the classical hamiltonian
$H$ is non extensive for $0 \leq \alpha \leq d$ and extensive for
$\alpha/d>1$. Through a $\tilde{N}$ rescaling of its potential energy
it can artificially be made extensive, thus merging with a long
standing tradition in mean-field-like approaches. Above
some critical value for the (rescaled) total energy per particle, the
weakening of the mixing properties is universal; more precisely, the
maximal Lyapunov exponents scale as $N^{-\kappa}$, $\kappa$ being a
function {\it only} of the ratio $\alpha/d$ (and not $d$ nor $U$). In
fact, it seems reasonable to expect that this $\kappa(\alpha/d)$
function is model-independent. More precisely, for total energies
high enough, it is expected to have one and the same functional form
$\kappa(\alpha/d)$ for {\it all} classical models whose potential is
nowhere singular and whose attractive tail decays as $1/r^{\alpha}$.
Consistently with this speculation, some specific gravitational and
fluid models \cite{GS} do exhibit $\kappa(0)=1/3$, in agreement with
the present results for the $XY$ models under analysis. An universal
$\kappa(\alpha/d)$ fits in fact very well the generic expectation for
such models within nonextensive statistical mechanics. Indeed, the
available studies of hamiltonian systems including long range
interactions  provide a variety of indications that a crossover
occurs, as a function of time, in the thermostatistical properties of
the system. For times below a crossover time $\tau(N)$, several
properties are anomalous (diffusion, velocity distribution and
others), whereas they become normal for times much larger than
$\tau$. The fact that $\tau$ seems to diverge with $N$ makes the
anomalous regime to become the only observable one in the
thermodynamic limit. It is precisely this regime which is focused on
within that generalized formalism. For the specific case of the
mixing properties, the present results suggest that a
less-than-exponential mixing takes place until a crossover time after
which the mixing becomes exponential, i.e., the usual one. This
crossover time might well scale like $1/\lambda_{max}'$ (or a
positive power of it). Therefore, since $\lambda_{max}'$ vanishes in
the limit $N \rightarrow \infty$, we expect the crossover time to
diverge, thus emerging the anomalous regime we were just discussing.



\begin{thebibliography}{99}
\bibitem{note1}
Definition (\ref{lambdat}) assumes an exponential separation. It is our
opinion that a diverging Lyapunov exponent indicates that the rate of
separation in these cases is more than exponential.
\bibitem{note2}
This rescaling implies that the microscopic coupling between rotators is
a function of the size $N$ of the system which is a macroscopic quantity.
\end{thebibliography}
\end{document}